\documentstyle[mprocl]{article}

\begin{document}

\title{Black holes of constant curvature}

\author{M\'aximo Ba\~nados }

\address{Departamento de F\'{\i}sica, Universidad de
Santiago de Chile, Casilla 307, Santiago 2, Chile, and \\ 
Centro de Estudios Cient\'{\i}ficos de Santiago, Casilla 16443,
Santiago, Chile}

\maketitle\abstracts{Black holes of constant curvature are constructed by identifying points in
anti-de Sitter space. In $n$ dimensions the resulting topology is
$\Re^{n-1} \times S_1$, as opposed to the usual $\Re^2 \times S_{n-2}$
Schwarzschild black hole.}

The goal of this talk is to report the existence of a family of black holes with constant curvature having the topology $\Re^{n-1} \times S_1$, as opposed to the usual $\Re^2 \times S_{n-2}$. These black holes 
were first discussed in Ref. [1] in four dimensions, although the higher dimensional nature of the causal structure was not exploited in that Reference.  The $\Re^{n-1} \times S_1$ exists in any dimension and can be regarded as a natural extension of the 2+1 black hole [2].  Here we shall only make a brief derivation of the solution. More details can be found in Ref. [3]. See also Ref. [4] for related work.  

In $n$ dimensions anti-de Sitter space is defined as the universal covering of the surface,
\begin{equation}
-x_0^2 + x^2_1 + \cdots + x^2_{n-2} + x_{n-1}^2 - x_n^2 = -l^2.
\label{ads/n}
\end{equation}
Consider the boost $\xi =(r_+/l) (x_{n-1} \partial_{n} + x_n
\partial_{n-1})$ with
norm $\xi^2=(r_+^2/l^2)(-x_{n-1}^2+x_n^2)$. We plot
parametrically the surface (\ref{ads/n}) in terms of the values of
$\xi^2$. There are two important values of $\xi^2$. First, for $\xi^2=r_+^2$ one has the null surface,
\begin{equation}
x_0^2=x_1^2 + \cdots + x_{n-2}^2,
\label{h/n}
\end{equation}
while for $\xi^2=0$ one has the hyperboloid,
\begin{equation}
x_0^2=x_1^2 + \cdots + x_{n-2}^2 + l^2.
\label{s/n}
\end{equation}

Let us now identify points along the orbit of
$\xi$. The region behind the
hyperboloid ($\xi^2<0$) has to be removed from the physical spacetime
because it contains closed timelike curves. The hyperboloid is thus a
singularity because timelike geodesics end there.  On the other hand, the null surface (\ref{h/n}) acts as a horizon because any physical observer that crosses it cannot go back. Indeed, the surface (\ref{h/n}) coincides with the boundary of the causal past of light like infinity.  
In this sense, the surface (\ref{ads/n}) with identified points represents a black hole. 

Let us now introduce local coordinates on anti-de Sitter space (in the region $\xi^2>0$) adapted to
the Killing vector $\xi$. We introduce the
$n$ dimensionless local coordinates $(y_\alpha,\phi)$ by, 
\begin{eqnarray}
x_\alpha &=& \frac{2l y_\alpha}{1-y^2}, \mbox{\hspace{1cm} }
\alpha=0,...,n-2
\label{y} \nonumber\\
x_{n-1}  &=& \frac{lr}{r_+}  \sinh\left(\frac{r_+\phi}{l}\right),
\nonumber \\ 
    x_n  &=& \frac{lr}{r_+}  \cosh\left(\frac{r_+\phi}{l}\right),
    \nonumber 
\end{eqnarray}
with $r = r_+ (1+y^2)/(1-y^2)$ and $y^2 = \eta_{\alpha\beta}\, y^\alpha y^\beta $ [$\eta_{\alpha\beta}=\mbox{diag}(-1,1,...,1)$]. The
coordinate ranges are $-\infty < \phi < \infty$ and $-\infty < y^\alpha <\infty$ with the restriction $-1<y^2<1$.

The induced metric has the Kruskal form,
\begin{equation} 
ds^2 =  \frac{l^2(r+r_+)^2}{r_+^2}\, dy^\alpha
dy^\beta\eta_{\alpha\beta} + r^2 d\phi^2, 
\label{ds/krus}
\end{equation}
and the Killing vector reads $\xi =\partial_\phi$ with
$\xi^2=r^2$. The
quotient space is thus simply obtained by identifying $\phi \sim \phi+2\pi n$, and the resulting topology is $\Re^{n-1}\times S_1$. The metric (\ref{ds/krus}) represents the $\Re^{n-1}\times S_1$
black hole written in Kruskal coordinates.  
Note that the above metric is a natural generalization of the
2+1 black hole. Indeed, setting $n=3$ in (\ref{ds/krus}) gives the 
non-rotating 2+1 black hole metric written in Kruskal coordinates
\cite{BHTZ}.

In five dimensions, let us introduce local ``spherical" coordinates
($t,\theta,\chi,r$) in the hyperplane $y^\alpha$:
\begin{eqnarray}
y_0 = f \cos{\theta} \sinh{(r_+t/l)}, &\mbox{\hspace{.6cm} }& 
           y_2 = f \sin{\theta} \sin{\chi},   \nonumber\\    
y_1 = f \cos{\theta} \cosh{(r_+t/l)}, &\mbox{\hspace{.6cm} }&
        y_3 = f \sin{\theta} \cos{\chi} \label{cc},
                      \nonumber
\end{eqnarray}
with $f(r) = [(r-r_+)/(r+r_+)]^{1/2}$. The metric (\ref{ds/krus}) acquires the Schwarzschild form,  
\begin{equation}
ds^2 = l^2 N^2 d\Omega_3 +  N^{-2} dr^2 + r^2 d\phi^2, 
\label{ds/sch}
\end{equation}
with $N^2(r) = (r^2 -r_+^2)/l^2 $ and
\begin{equation}
d\Omega_3 = - \cos^2{\theta}\, dt^2 + \frac{l^2}{r_+^2} (d\theta^2 +
\sin^2{\theta} d\chi^2). 
\label{sph}
\end{equation}
The horizon in these coordinates is located at $r=r_+$, the point
where $N^2$ vanishes.  Note, however, that these coordinates are meaningful only in the exterior region and they cannot be extended to $r<r_+$.  Also, the ``static" coordinates do not cover the full outer manifold. Actually, it can be proved that there is no globally defined time like Killing vector on this geometry (see Holst and Peld\'an in Ref. [4]).

The black hole just constructed has an Euclidean sector which can be
obtained by setting $\tau = it$ in (\ref{ds/sch}), or $y_0 \rightarrow
iy_0$ in (\ref{ds/krus}). In the Euclidean sector, the ``static" coordinates do cover the full Euclidean manifold.  

Just as in 2+1 dimensions, angular momentum in the plane $t|\phi$ can be
added by considering a different Killing vector to do the identifications. This is most easily done by setting $r_+=l$ in (\ref{ds/sch}),
making the replacements,
\begin{eqnarray}
t   &\rightarrow& \beta t\frac{r_+}{l^2} +  (\phi - \Omega \, \beta t)
\frac{r_-}{l},
\\
\phi &\rightarrow& \beta t\frac{r_-}{l^2} + (\phi - \Omega \,
\beta t) \frac{r_+}{l},
\label{ang}
\end{eqnarray}
($r_+>r_-$ arbitrary constants), and identifying points along the new
angular coordinate $\phi$: $\phi \sim \phi + 2\pi n$. The constant $r_+$
parametrizes the location of the outer horizon, and the new metric
has two independent conserved charges. In the Euclidean
formalism, the time coordinate $\tau=-it$ must be periodic in order to
avoid conical singularities. This gives the value $\beta = (2\pi
r_+l^2)/(r_+^2-r_-^2)$ [with $0\leq t <1$] which can be interpreted as the inverse temperature of the black hole.

Since the above geometries are locally anti-de Sitter, they are natural
solutions of Einstein equations with a negative cosmological constant.
However, due to the non-standard asymptotic behaviour of (\ref{ds/sch}) one finds that all conserved charges are infinite.  Global charges associated to these black holes can be defined in the context of a Chern-Simons supergravity theory in five dimensions proposed sometime ago by Chamseddine \cite{Chamseddine}. This action is constructed as a Chern-Simons theory for the supergroup $SU(2,2|N)$ \cite{Chamseddine}. The energy $M$ and angular momentum $J$ of the black hole embedded in this supergravity theory are,
\begin{equation} 
M = \frac{2r_+ r_-}{l^2}, \ \ \ \      
J = \frac{r_+^2 + r_-^2}{l}.
\label{J}
\end{equation}
The entropy on the other hand is equal to,
\begin{equation}
S =  4\pi \, r_-.
\label{S}
\end{equation}      
This result is rather surprising because it does not give an entropy
proportional to the area of $S_1$ ($2\pi r_+$).  A similar phenomena has been reported by Carlip et al \cite{Carlip-Gegemberg-Mann}. The entropy given in (\ref{S}) satisfies the first law, 
\begin{equation}
\delta M = T \delta S + \Omega \delta J, 
\label{fl}
\end{equation}
where $M$ and $J$ are given in (\ref{J}) and $T=1/\beta$. 

During this work I have benefited from many discussions with Andy Gomberoff, Marc Henneaux, Cristi\'an Mart\'{\i}nez, Claudio Teitelboim and Jorge Zanelli.  I would also like to thank Peter Peld\'an for many remarks and useful suggestions, and Dieter Brill for comments on a previous version of the manuscript. This work was partially supported by the grant \# 1970150 from FONDECYT (Chile), and institutional support by a group of Chilean
companies (EMPRESAS CMPC, CGE, COPEC, CODELCO, MINERA LA ESCONDIDA,
NOVAGAS, ENERSIS, BUSINESS DESIGN ASS. and XEROX Chile).

\end{document}